# Nanotechnology and Processes
# The Nano-Photovoltaic Panels

## Manuel Alberto M. Ferreira

Instituto Universitário de Lisboa (ISCTE-IUL), Portugal

manuel.ferreira@iscte.pt

## José António Filipe

Instituto Universitário de Lisboa (ISCTE-IUL), Portugal

jose.filipe@iscte.pt

## José Chavaglia Neto

Instituto Universitário de Lisboa (ISCTE-IUL), Portugal

jnchavaglia@gmail.com

## Abstract

Nanotechnology may work as a powerful weapon to be used for creating competitive advantages in the energy market. Using the photovoltaic nano-panels, which may reduce considerably the production costs and meet simultaneously socio-environmental requirements demanded by law. It is a way to produce clean energy in innovative terms. Moreover, today the adoption of nanotechnology in energy production can make this kind of energy very interesting along the years. Nanotechnology may be responsible for considerable gains, both economically and the ones resulting from its contribution to protect the planet against pollution.

## 1. Introduction

In this new era in which competition became global, technological innovations are the way for companies overcoming the challenges in the context of competition, by applying innovative processes to their production structures.

Presently, there is a permanent run to technological innovations by companies in the market aiming to go ahead in the competition struggle. This has been happening especially in the last decade in which the life cycle of products has decreased considerably. Many developments in terms of innovations have been seen in sectors as communications, medicine, robotics, computing, energy, or many other areas of society.

It is very relevant to see the role that internet took over, the development of mobile devices, new medical imaging, the self-service centers, banks, etc. The truth is that people have become dependent on new technologies and companies became dependent on the process of constant search for innovation. As some people do better professionally than others, some companies also do better in the competitive environment than others. Notably, in both cases, both individuals and companies that stand out have something in common: the ability to innovate.

In this scenario, this paper aims to highlight one very significant scientific advance in human history - the nanotechnology, which opens unimaginable possibilities in various fields of human reality and in various fields of science. Today the influence of nanotechnology can be seen in areas as medicine (implantation of nano-robots, for example), agriculture (for pest control) or many others. The possibilities are endless on these fields.

This work shows the use of nanotechnology in the industry, discussing how the use of nano-panels may provoke very interesting gains in the industry, by innovating in a topic that permits long term gains in several domains. The use of nanotechnology in the industrial production processes is evidenced in the electric energy produced by photovoltaic panels. Emphasis is placed on the competitive advantage associated with the use of nanotechnology to solar energy production for companies in this market segment.

The American firm "Nanosolar", which studies are sponsored by major companies like Google or IBM or resulting from the allocation of benefits offered by the Department of Energy, is leading the race for energy production derived from nanotechnology. It has named this technology as "nano-photovoltaic panels".

The use of this product has showed a reduction in total costs on firms in relation to other types of solar energy and meets government requirements and the importance of social use of energy when it is obtained from clean sources.

Despite the high feasibility for the economy and the environment, there are some considerations regarding the ethical and moral limits on nanotechnology that should be considered.

## 2. The Emergence of Nanotechnology

The first person to conceptualize Nanotechnology was Richard P. Feynman, although he had not used this term in his speech to the American Physical Society on December 29, 1959, where he made the first comments on the subject. However, the word "nanotechnology" was first used by Professor Norio Taniguchi (1974) to define the fabrication of a scale of 1 nm. Nanotechnology is the potential ability to create things from the smallest element, using the techniques and tools that are being developed today to place every atom and molecule in place. The use of nanometer implies the existence of a system of molecular engineering, which will likely generate the revolution of the factory manufacturing model as it is known.

Nanotechnology will offer, in addition to products of higher quality at lower cost, a range of possibilities to generate new means of production and new types of resources and factors. This is a manufacturing system that could produce more manufacturing systems (plants that produce other plants) in a quick, cheap, and clean way. The means of production may be reproduced exponentially. So, in just a few weeks, power would pass from a few to several billion nano factories. Thus represents a kind of revolutionary technology, manufacturing, powerful, but also with many potential risks, besides the existing and well recognized benefits (see Euroresidentes, 2011, p. 01).

In Brazil, the budget of the Ministry of Science and Technology for the next four years is R680 million. Overall, it is estimated that only developed countries to allocate a sum of around USD 5.5 billion. An important example of successful application of nanotechnology is *Empresa Brasileira de Agropecuária* (Embrapa). It has been working with nanotechnology in various research centers and has already released some products. One of the most notable is perhaps the "electronic tongue", a device that combines chemical sensors with nanometer-thick with a computer program that detects flavors and aromas and serve to quality control and certification of wines, juices, coffees, and other products (see DIEESE, 2008, p. 03).

There is an interesting reconfiguration of the industrial model, directly related to the use of nanotechnology in the various branches of economic activities. This study gives an outline for the manufacture of electric power, which generically McKibben considered itself as a country's economy (see McKibben, 2009, p. 24).

Nanotechnology is therefore contributing for the transformation of traditional models, independently either of the way goods and services are produced or of the way the production is conducted and made.

### 3. Nano-photovoltaic panels as an innovation

Nowadays the solar panels are a way to produce energy. There is a well-known level of efficiency and expectations that are got with the traditional panels, which may be compared with new forms of producing energy when using other kind of non-traditional panels.

Considering that, some notes about innovation and new forms of obtaining efficient processes get required. Companies need to be dynamic in the development of innovations and thus creating competitive advantages through its production processes so they can create economic value, and consequently generate their viability based on the market in which they operate. For McDonough III (2009, p. 04) "in the current buoyant economy, organizations must continually reinvent what they are and what they do [...]". This means that they need to constantly maintain market differentiation, through deliberate strategies to obtain competitive advantages that provide monopoly profits, even if temporary in this environment that requires from companies a high degree of competitiveness.

The competition is part of a dynamic and evolutionary operation of the capitalist economy. The evolution of this economy is seen as over time based on an uninterrupted process of introduction and diffusion of innovations in a broad sense, i.e., any changes in the economic space in which these companies operate, whether changes in products, processes, sources of raw materials, forms of productive organization, or in their own markets, including in terms of geography (see Schumpeter, Brazilian version, 1982, p. 65).

According to McAfee and Brynjolfsson (2008, p. 78) "the result is that an innovator with a better way of doing things can grow at unprecedented speeds and dominate the industry".

In contrast, at the same time, which seeks a high level of competitiveness, it is understood that there is a need for companies to retain the existing common resources for their optimization in the future because it is not possible no longer count on such a supply of natural resources to meet the continuing huge demand, given the level of production that humanity has achieved over the last century.

To Nogami and Passos (1999, p. 03), the harsh reality of scarcity arises the necessity of choice. Since it is not possible to produce everything that people want, mechanisms must be created to somehow show the societies the path to decide what goods to produce and which needs are met.

The appropriated concepts which are consistent with the possibility of economic efficiency and technological means demand a study that requires an integrated analysis of the proceedings related to the production combination of forces and the structure of supply inside the production unit in order to get efficient processes. All the means or methods of production indicate some of these combinations. Production methods vary in the way how such combinations occur, or by objects or by the combined ratio of their quantities. Every concrete act of production incorporates some combination thereof. It could also be considered as a combination enterprise itself, and even the production conditions of the whole economic system (see on this subject Schumpeter, 1982, p. 16).

The possibility of economic and technological efficiency reflects a producing combination of forces and inputs to reach an interesting production level for the company. All the means or methods of production indicate different sorts of combinations. Production methods vary in the way by which such combinations occur, or by objects or by the combined ratio of their quantities. Every concrete act of production incorporates a kind of combination. A company can itself be considered a combination by itself, and even the production conditions of the whole economic system (see Schumpeter, 1982, p. 16).

Thus, it can be said that any company when producing goods and services using clean energy (in particular, the use of photovoltaic panels produced by nano panels) generates competitive advantage by breaking the closed circle of the economy. This is made by creating a new mechanism of generation of market value, since it is a new way to produce through a new combination of available resources.

In a direct way, besides the fact that the use of photovoltaics is already an innovation on itself, when it is combined with nanotechnology, its power can also be interpreted as the ability to break paradigms in the energy industry, taking the form of a powerful competitive weapon of production units. Given this framework, nanotechnology combined with solar energy production becomes evidently interesting for the production units, whether public or private.

## 4. The Utilization of Nano panels in the industry

Considering Chavaglia and Filipe's recent research for solar panels in Brazilian market, some results are presented in terms of the nano panels technology and some specific examples for Brazil are shown.

To describe the way how a nano panel is got, it can be said that first it is necessary to produce the semiconductor's nanoparticles (about 20mn in size, equivalent to 200 atoms in diameter). Then, aluminum sheets are placed in press, similar to those used in graphic paper. These aluminum sheets may be very dynamic in their use, because of their length and their width. This makes the product much more adaptable to formats required for the product. Then a thin layer of semiconducting ink is painted on the aluminum substrate. After that, another press put layers of cadmium sulfide and sulfur, and zinc oxide (CdS and ZnO). The layer of zinc oxide is non-reflective to ensure that sunlight can reach the semiconductor layer. Finally, the sheet is defined in sheets of solar cells. Unlike other methods of panels manufacture that are usually used, which typically requires a special location for manufacturing, nano-panels can be produced outdoors (Chavaglia *et al*, 2012).

This technology is very interesting, reaching good results either in terms of costs when compared with other solar panels and in terms of performance even considering public energy. Those costs have coming to be considerably reduced.

In Brazil a strong trend of increasing costs to the production of hydroelectric power is observed. The comparison of the hydroelectric energy produced in Brazil and other countries like Canada, for example, shows the importance of the development of alternative energies. In Brazil, the behavior of the total cost of producing energy used to be well above inflation in this country for a long period. Recently, Brazil took the advantage of using photovoltaic energy using nano-panels. It may be noted in fact that this is one of the reasons that qualify the use of photovoltaic energy as a generator of competitive advantage in the market. The company which owns such technology may experience a reduction on its variable costs, when compared with solar photovoltaic panels (silicon). Nano-panels permit also to perceive a competitive advantage in environmental conservation terms, if compared with the energy provided by public network (see Chavaglia *et al*, 2012).

Therefore, nanotechnology is contributing for the transformation of traditional models. This transformation may be in the way goods and services are produced, or in the way the production is conducted and made.

## 5. Final Remarks

The possibilities open to mankind through nanotechnology are enormous. It encompasses the way people connect with each other and with the natural resources to produce goods and services.

First, it is prudent to recognize that there are dangers when using nanotechnology. These dangers may get materialized under the prism of problems arising from human nature itself, as corruption and bad faith, or the overexploitation of resources.

However, the advantages arising from the use of nanotechnology may mean breaking with paradigms in what concerns to the attendance of the consumption needs of humanity. These advantages may be visible in particular in terms of reducing costs and increasing the quality of both the products and the production method and can be seen in the case of electrical energy production based on nano-photovoltaic panels.

Therefore, it is necessary to be precautious when using nanotechnology It is important to stress that the benefits can be got not only for the production units as well as for society in a whole. Its use allows mankind to gain additional levels of wealth and quality of life. Anyway, the balance between costs and benefits must be taken into consideration in the analysis of all the relevant elements.

The experience shows that nano-photovoltaic panels may bring interesting results for economies and bring a new form of producing energy with considerable advantages in the long term.